\documentclass[english,prl,aps,twocolumn]{revtex4-1}
\usepackage{amsmath}
\usepackage{amssymb}
\usepackage{graphicx}
\usepackage{float}
\usepackage{amsthm}
\usepackage{physics}%

\begin{document}

\title{Transfer arbitrary photon state along a cavity array without initialization}

\author{Yang Liu}

\affiliation{Beijing National Laboratory for Condensed Matter Physics,
  and Institute of Physics, Chinese Academy of Sciences, Beijing
  100190, China}

\author{D. L. Zhou}

\affiliation{Beijing National Laboratory for Condensed Matter Physics,
  and Institute of Physics, Chinese Academy of Sciences, Beijing
  100190, China}

\email{zhoudl72@iphy.ac.cn}

\begin{abstract}
  We propose a quantum state transfer (QST) scheme that transfers any
  single-mode photon state along a one-dimensional coupled-cavity
  array (CCA). By building a map from QST in a CCA to that in a
  spin-$\frac{1}{2}$ chain, we show that all the previous results of
  QST schemes for the spin chain system find paralleled applications
  in that in the CCA system. Further more, high fidelity QST along a
  long CCA can be achieved for arbitrary initial states. Using numerical
  simulations we provide a visual presentation of the result: at some
  time $\tau$ the CCA system get high fidelity QST under different
  initial conditions. Finally we discuss possible experimental
  realizations of our QST scheme.
\end{abstract}

\pacs{03.67.Ac, 03.65.-w}

\maketitle

\paragraph*{Introduction.}---
Quantum state is the carrier of the information in quantum information
and quantum computation. Transmitting quantum state from one location to
another is one of the basic tasks in quantum information processing
system. The most famous scheme of QST is quantum state
teleportation \cite{PhysRevLett.70.1895}, where the unknown state is
teleported with the aid of one shared EPR pair between the sender and
the receiver and 2 bits of classical information. This scheme indicates
that quantum entanglement is a resource in QST. A more direct one is
to transfer the unknown state through a shared quantum
network~\cite{PhysRevLett.78.3221, PhysRevLett.91.207901}.

The simplest quantum network used to transfer quantum state is a one
dimensional spin-$1/2$ spin chain, which is pioneered by Bose. Bose
showed that the high fidelity of state transfer could be achieved
through a long unmodulated spin chain. QST along an unmodulated spin
chain can be perfect only when the length of the spin chain is less
than $4$. For the chain of any length perfect QST can be achieved by
modulating the coupling strengths between adjacent spins
\cite{PhysRevLett.92.187902, PhysRevA.71.032309, 0295-5075-65-3-297,
  0953-8984-16-28-019}. Other schemes are also discussed such as only
tuning the two end coupling strengths to get high fidelity QST
\cite{PhysRevA.72.034303, PhysRevA.76.052328, PhysRevA.71.022301,
  1367-2630-12-2-025019, PhysRevA.78.022325, PhysRevLett.106.040505,
  PhysRevA.85.022312}, QST without initialization
\cite{PhysRevLett.101.230502, PhysRevA.79.054304}, and generalizing to
the high spin QST \cite{Bayat2007,Qin2013}. Number-Theoretic relation
between QST and the length of one-dimension spin chain is found in
Ref.~\cite{PhysRevLett.109.050502}.

In addition, schemes based on cavity quantum electrodynamics are also reported.
An initial proposal is to transfer the state of a qubit from a
cavity-atom system to another one through an optical fiber connecting
the two cavities \cite{PhysRevLett.78.3221, Ritter2012}.

In this paper we propose a QST scheme that transfers any single-mode
photon state along a one-dimensional CCA. All the previous results got
in the spin chain system mentioned above are applicable in our scheme
and the initialization step is not needed. It is
naturally a high dimension QST scheme. With the development of
technology of producing high quality cavities \cite{Armani2003,
  Yariv:99, Akahane2003} and the control of the photons in the cavity
\cite{Kuhr2007, Wang2008, Brune2008, Johnson2010a, Sayrin2011}, the
realization of our scheme is possible. 

This article is organized as follows. First we propose the QST scheme,
where the Hamiltonian of the system is given. Next we analyse the
fidelity of QST in our scheme and give the condition of perfect
QST. Then we solve the dynamic problem about fidelity. After that we
simulate the QST using our scheme in three cases: uniform coupling
CCA, perfect modulated CCA and the CCA with coupling strengths in the
ballistic regime. Finally we give some discussion on experimental
realization of our scheme.

\begin{figure}[htbp]
  \centering
  \includegraphics{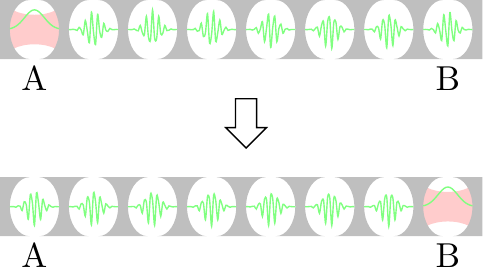}
  \caption{Our quantum communication protocol. Initially prepare the
    quantum state needed to communicate in the first cavity (cavity
    A). After a while, the state transfer to the other end of the
    array (cavity B). The Gauss curve means the transmitted photon
    state. The wavy lines in the cavities represent the arbitrary
    state of single-mode photon. \label{fig:Transfer-a-photon}}
\end{figure}

\paragraph*{Scheme and Analysis.}--- The system of our scheme is a CCA
as depicted in Fig.~\ref{fig:Transfer-a-photon}. Every cavity has the
same cavity mode $\omega$. Photons can hop between adjacent cavities
due to the overlap of the light mode \cite{Hartmann2006}. The
Hamiltonian is given by
\begin{equation*}
  H = \hbar \omega \sum_{n=1}^{N}\hat{a}_{n}^{\dagger}\hat{a}_{n} + \sum_{n=1}^{N-1}
  J_{n} (\hat{a}_{n}^{\dagger}\hat{a}_{n+1}+\hat{a}_{n}\hat{a}_{n+1}^{\dagger}),
\end{equation*}
where $\omega$ is the frequency of the cavity mode, $J_{n}$s are the
coupling strengths between adjacent cavities, which can be adjusted by
changing the thickness of the mirrors.

The process of the QST along the CCA is as follows. First, the state
we want to transfer is encoded on the photons in the first cavity
(cavity $A$) as $|\phi_{1}\rangle=f(\hat{a}^{\dagger}_1)|0\rangle$
which is unknown in many cases. Next we allow the unitary evolution
controlled by the Hamiltonian $H$ for a time period $t$. Then we check
whether the unknown state has been transferred to another end of the
array (cavity $B$ ($N$)).

Firstly we consider the fidelity which is defined as
$\langle\phi_N|\rho_N(t)|\phi_N\rangle$ to characterize the quality of
the QST. Let the initial state of the system as
\begin{align*}
  |\psi_{0}\rangle\langle\psi_{0}| &
  =f(\hat{a}_{1}^{\dagger})|0\rangle\langle0|f^{*}(\hat{a}_{1})\otimes\rho_{2-N}.
\end{align*}
Then the fidelity of the system at time $t$ is
\begin{align*}
  \mathcal{F}(t)
  =&\tr\left(\langle0|f^{*}(\hat{a}_{N}(t))f(\hat{a}_{1}^{\dagger})|0\rangle\langle0|f^{*}(\hat{a}_{1})\right.\\
  &\ \ \ \
  \left.\otimes\rho_{2-N}f(\hat{a}_{N}^{\dagger}(t))|0\rangle\right),
\end{align*}
where $\hat{a}_N(t)$ is operator $\hat{a}_N$ in the Heisenberg
picture, that is $\hat{a}_N(t)=\hat{U}^\dagger(t) \hat{a}_N\hat{U}(t)$
with $\hat{U}(t)$ as time evolution operator. If at time $\tau$ we
have the relation $\hat{a}_{N}\left(\tau\right)=\hat{a}_{1}$, then we
get the conclusion that at time $\tau$ we have a perfect transfer,
$\mathcal{F}(\tau)=1$.
In other words, to get a perfect photon state transfer means to get a time
$\tau$ that $\hat{a}_{N}(\tau)=\hat{a}_{1}$. It can be easily verified
by noting that the expect value of the any operator in the $N$-th
cavity at time $\tau$ is equal to that of the operator in the first
cavity at initial state, e.g.,
$\langle\hat{a}_N^\dagger\hat{a}_N(\tau)\rangle=\langle\hat{a}_1^\dagger\hat{a}_1(0)\rangle$.

Now we analyse the dynamics of $\hat{a}_{N}^{\dagger}(t)$, which satisfies
the Heisenberg equation
\begin{equation}
  \frac{d \hat{a}_{N}^{\dagger}\left(t\right)}{dt} = i \left[H,\,
    \hat{a}_{N}^{\dagger}(t) \right].
\end{equation}

First we note that the set,
$\{\hat{a}_{n}^{\dagger}\vert{n}=1,\,2,\,3,\cdots,\,N\}$, is closed
under the action $[H,\cdot]$. So $\hat{a}_{N}^{\dagger}(t)$ can be
expanded as
\begin{equation}
  \hat{a}_{N}^{\dagger}\left(t\right) =\sum_{n=1}^{N}\alpha_{n}(t)\hat{a}_{N+1-n}^{\dagger}.
\end{equation}
Now we come to the solution of $\hat{a}_{N}^{\dagger}(t)$, which is
determined from the Heisenberg equation for $\hat{a}_{N}^{\dagger}$:
\begin{equation}
  \frac{dA}{dt}=i(G+\hbar\omega)A, \label{eq:evolution_equation}
\end{equation}
where $A=[\alpha_{1}(t), \, \alpha_{2}(t), \,\cdots,\alpha_{N}]^{T}$
with $T$ being the transpose operation, $G$ is a tri-diagonal matrix
\[
G=\begin{bmatrix}0 & J_{N-1} & 0 & 0 & \cdots \\
  J_{N-1} & 0 & J_{N-2} & 0 & \cdots \\
  0 & J_{N-2} & 0 & J_{N-3} & \cdots \\
  0 & 0 & J_{N-3} & 0 & \cdots \\
  \vdots & \vdots & \vdots & \vdots & \vdots
\end{bmatrix}.
\]
The initial condition is $A(0)=[1,\,0,\,\cdots,0]^{T}$. To solve the
differential equations we can apply the Laplace transformation on the
both sides of equation as it was done in Ref.~\cite{Liu2013}. Note
that multiplying the both sides of Eq.~\eqref{eq:evolution_equation}
by $i$, we can rewrite it as $i\frac{dA}{dt}=\hat{H}_{new}A$, which
has the same formation as Schr\"odinger equation with
$\hat{H}_{new}=-(G+\hbar\omega)$. The new Hamiltonian $\hat{H}_{new}$
is in an $N$-dimensional Hilbert space, which is much more tractable
than the original Hamiltonian $\hat{H}$ that is in the
$D^{N}$-dimensional Hilbert space. $A$ is the wave function of the new
Hamiltonian, and we denote it as $|A\rangle$. That is the operator
$a^\dagger_N(t)$ is represent as a vector $|A\rangle$ in the Hilbert
space of the new Hamiltonian. It is worth mentioning that the reason
of the less Hilbert space is that the number of the set, which
contains $A_{N}^{\dagger}$ and is closed under the operator
$[H,\cdot]$, is only $N$, rather than the excitation number
conservation. This can be seen clearly in the XY Hamiltonian with the
coupling strength that can't conserve the excitation number
\cite{Liu2013}.

In the uniform condition, $ J_{n}=1 $, the eigenvalues and eigenstates
of the new Hamiltonian $\hat{H}_{new} $ are $ E_{n}= -2\cos\frac{\pi
  n}{N+1}- \hbar\omega $, with $ n=1,\cdots,N $, and $\langle
l|\phi_{n}\rangle=\sqrt{\frac{2}{N+1}}\sin\frac{\pi nl}{N+1}.$ We
consider the question that what is the sate of the new system at the
given time $ t $. The Hamiltonian of the system is $ \hat{H}_{new} $
and the initial state is $|A_{0}\rangle=[1,0,\cdots,0]^{T} $. Using
the Schr\"odinger equation we know
\[
|A(t)\rangle
=\sum_{n}\exp(-iE_{n}t)|\phi_{n}\rangle\sqrt{\frac{2}{N+1}}\sin\frac{\pi
  n}{N+1}.
\]
The last element of the state is
\[
\alpha_N(t) =
\sum_{n}(-1)^{n-1}\frac{2}{N+1}\exp(-iE_{n}t)\sin^{2}\frac{\pi n}{N+1}.
\]

From  Ref.~\cite{PhysRevLett.109.050502} we know that if and only
if the number of length is $ N=p-1,$ $2p-1 $, where $ p$ is a prime,
or $ N=2^{m}-1$ (for convenience we call it pretty good length
condition), there is a time $\tau $ that $
\exp(-i\mathcal{E}_{n}t)\thickapprox(-1)^{n-1}\gamma,$ where
$\gamma=1$ if $N\equiv1\mod4$, $\gamma=-1$ if $N\equiv3\mod4$,
$\gamma=\pm i$ if $N$ is even, and $
\mathcal{E}_{n}=E_{n}+\hbar\omega$.

So we have $ \alpha_{N}(\tau) \thickapprox \gamma
e^{i\hbar\omega\tau}$. From $|\gamma e^{i\hbar\omega\tau}|=1,$ and the
normalization of the state we get that
$\alpha_{1}(\tau)\thickapprox\alpha_{2}\left(\tau\right)\thickapprox\cdots\thickapprox\alpha_{N-1}(\tau)\thickapprox0$. So
at the time $\tau$ we have
\[
\hat{a}_{N}^{\dagger}\left(\tau\right) \thickapprox \gamma e^{i\hbar\omega\tau}\hat{a}_{1}^{\dagger}.
\]
As for the phase $ \gamma e^{i\hbar\omega\tau}$, we can adjust the
cavity mode to a proper value $ \omega_\tau$ that make
$\gamma{e}^{i\hbar\omega_{\tau}\tau}=1$. So the get the conclusion
that we get pretty good state transfer (PGST) at time $\tau$ if the
length of the cavities satisfies the pretty good length condition and
the cavity mode is $\omega_\tau$. Compared with the PGST in spin
chains we don't need the initialization of the cavities or the single
excitation condition. In XY spin chains system QST is proportional to
the parity of the initial state \cite{Liu2013}, while in the CCA
system if we can achieve perfect QST at time $\tau$ then the initial
state of cavities $ 2-N$ have nothing to do with QST at the perfect
time. The reason is that in XY spin chain system the operators in
first site is bound with the other part of system by $X_1(Y_1)Z_{2-N}$
while in the CCA system $\hat{a}_1^{\dagger} $ is standalone in the
Heisenberg equation related with the operators of the $N$-th site.

For the general case that $J_n$s are not uniform, $\alpha_N$ is
provided in ref. \cite{Liu2013} as
\begin{equation}
  \frac{\alpha_{N}(t)}{\det A_{N}^{(N)}} = 
  \begin{cases}
    \sum_{i=1}^{M} \frac{\sin(q_{i} t)}{q_{i} \prod_{j\neq i}
      (q_{j}^{2}-q_{i}^{2})} & \text{for } N=2m,\\
    \sum_{i=0}^{M} \frac{\cos(s_{i}t)}{\prod_{j\neq i}
      (s_{j}^{2}-s_{i}^{2})} & \text{for } N=2m+1,
  \end{cases}
\end{equation}
where $A=p-G$ with $p$ as Laplace complex argument, $A_{N}^{(N)}$ is
the matrix $A$ whose $N$-th column vector is replaced by $A(0)$. $q$
and $s$ are roots of $\det A_{N}^{(N)}$. $m$ is an integer. When
$\alpha_N(\tau)=1$ perfect QST is got.

\begin{figure}[htbp]
  \includegraphics{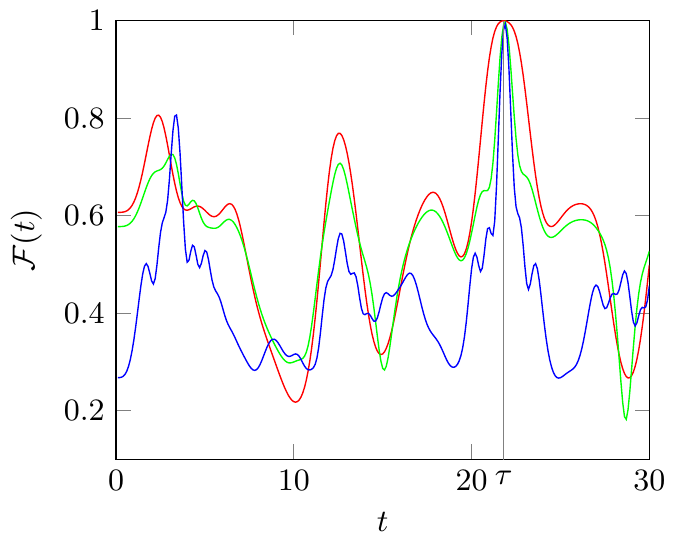}
  \caption{(Color online) Numerical simulation of the fidelity $F(t)$
    of QST for $N=5$ as a function of $t$ in uniform coupling case
    ($J_i=1$). The initial states of the three lines are
    $|\alpha\rangle|0000\rangle$, $\frac{1}{\sqrt{3}} (|0\rangle +
    |1\rangle + |2\rangle)|1000\rangle$, $\frac{1}{\sqrt{14}}
    (|0\rangle + 2|1\rangle +3|2\rangle) |1100\rangle$
    respectively. $|\alpha\rangle=e^{-|\alpha|^{2}/2}e^{\alpha
      a^{\dagger}}|0\rangle$ is coherent state, here
    $\alpha=1$.\label{fig1}}
\end{figure}

\paragraph*{Numerical simulation.}--- Now we numerically simulate the
QST in the unmodulated CCA and show its result in Fig.~\ref{fig1}. The
system we simulate has length $ N=5$ with coupling strength $J_n=1 $.
Here we choose units such that $\hbar=1$. It shows that at time $
\tau=21.8$, which requires $ \omega(k)=\frac{2k\pi}{\tau}$, $
k=0,1,2,\cdots$, we get a good fidelity $F(\tau)=0.9999$, for any
initial state of the chain $2-N$ and the sent state. The first line
(red one) simulates the fidelity of QST with the initial state of the
cavities $2-N$ being $|0000\rangle$ and the sent state being the
coherent state $|\alpha\rangle=e^{-|\alpha|^{2}/2}e^{\alpha
  a^{\dagger}}|0\rangle$, $\alpha=1$. The initial state of cavities
$2-N$ for the other two lines (green and blue ones) are
$|1000\rangle$, $|1100\rangle$ and the sent states are
$\frac{1}{\sqrt{3}}(|0\rangle +|1\rangle +|2\rangle)$ and
$\frac{1}{\sqrt{14}}(|0\rangle +|1\rangle +|2\rangle)$,
respectively. The $\omega$ we choose is $\omega(1)=0.288$.

Note that other conclusions of QST in the spin chain system are also
applicable in the CCA system. As we know that the spin chains with
modulated coupling strength have the prefect QST when the parity of
the initial state (except the first spin) is 1. The modulated coupling
strengths are $J_{n}=J_{n}^{[k]}=\sqrt{n(N-n)}$ for even $n$ and
$J_{n}=J_{n}^{[k]}=\sqrt{(n+2k)(N-n+2k)}$ for odd $n$, where
$k\in\{0,\,1,\,2,\cdots\}$
\cite{PhysRevLett.92.187902,PhysRevA.71.032309}. For the case $k=0$,
the matrix $G$ is identical to the representation of the Hamiltonian
$H$ of a fictitious spin $S=\frac{1}{2}(N-1)$ particle: $H=2S_{x}$,
where $S_{x}$ is angular momentum in $x$ direction
\cite{PhysRevLett.92.187902}.

\begin{figure}[htbp]
  \centering
  \includegraphics{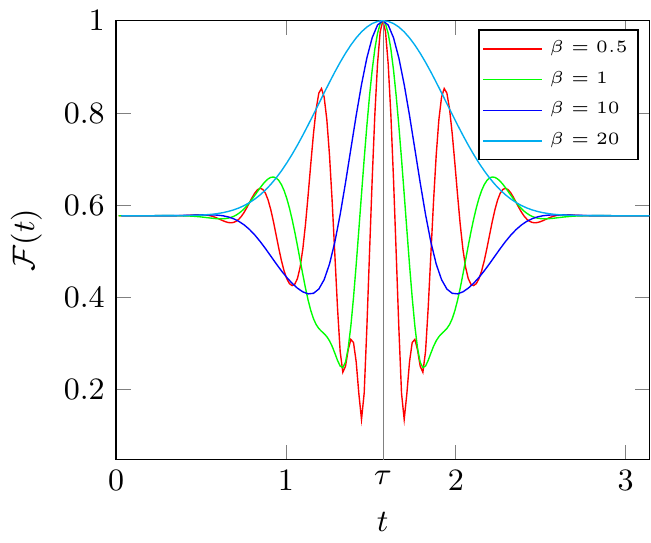}
  \caption{(Color online) Numerical simulation of the fidelity $F(t)$
    of QST for $N=8$ as a function of $t$ in the modulated coupling
    case ($J_n=\sqrt{(n+2k)(N-n+2k)}$) for different initial
    states. The initial states are
    $\rho_1\otimes\rho_{2-8}s$, and
    $\rho_1=\frac{1}{3}(|0\rangle+|1\rangle+|2\rangle)(\langle{0}|+\langle{1}|+\langle{2}|)$.
    \label{fig:fig2}}
\end{figure}

Now we consider the case that the coupling strengths are the prefect
modulated ones with $k=0$. So the new Hamiltonian is
$H_{new}=-(2S_{x}+\hbar\omega)$.  $\alpha_{N}(t)$ can be written
directly as
\[
\alpha_{N}\left(t\right)=\left[i\sin\left(t\right)\right]^{N-1}e^{i\hbar\omega t}.
\]
So at time $\tau=\frac{\pi}{2}$, $|\alpha_N(t)|=1$. The required
frequency is $\omega=4k+1-N$, $k=0,1,2,\cdots$. In Fig.~\ref{fig:fig2}
we demonstrate the fidelity $\mathcal{F}(t)$ versus $t$ for the
modulated CCA system with length $N=8$. The sent state is
$\frac{1}{\sqrt{3}} (|0\rangle + |1\rangle +|2\rangle ) $, and the
initial states of cavity $2-N$ are thermal state
$\frac{e^{-\beta{H}}}{Z}$ with $\beta=0.5,1,10,20$ and the
$\omega=17,9,5,1$ respectively. It shows that at time $\tau=\pi/2$ QST
of the CCA system with modulated coupling strength is perfect whatever
the initial state of cavities $2-N$ are. Fig.~\ref{fig:fig2} also
depicts that different frequencies, $\omega$s, result in the different
oscillation times in one period in the fidelity aspect as expected.

In the uniform XX spin channel, perfect QST can be achieved by tuning
down the two end coupling strengths limited to zero for arbitrary
length $N$~\cite{PhysRevLett.106.040505}. But the optimal time of
perfect QST becomes long as end coupling strengths decreasing. There
is a regime, which is called ballistic regime, that $0<J_{end}<1$ (
the uniform coupling strength is set to 1) the fidelity of QST is high
and the transmission time is $t \sim N$ \cite{1367-2630-13-12-123006}.

\begin{figure}[htbp]
  \centering
  \includegraphics{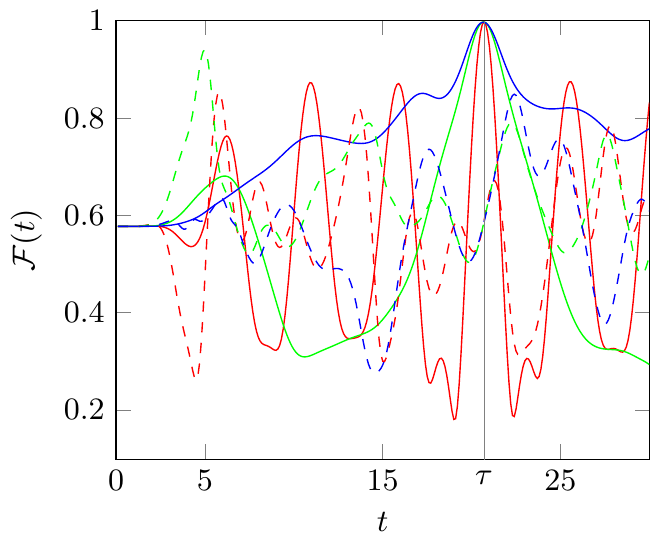}
  \caption{(Color online) Numerical simulation of the fidelity $F(t)$ of
    QST for $N=8$ as a function of $t$ in the ballistic regime case
    ($J_1=J_7=0.3$, $J_{n\neq 1,7}=1$) for different initial states and
    photon frequency $\omega$ (solid line). The initial states are
    $\rho_1\otimes\rho_{2-8}s$, 
    $\rho_1=\frac{1}{3}(|0\rangle+|1\rangle+|2\rangle)(\langle{0}|+\langle{1}|+\langle{2}|)$,
    $\rho_{2-8}$s are thermal state with $\beta=20$ (red line),
    $|1000000\rangle$ (green line), $|1100000\rangle$ (blue line),
    respectively. And the corresponding photon frequencies are $1.29$,
    $0.379$, $0.076$. The dashed lines depict the corresponding fidelity
    in the uniform coupling system. \label{fig:ballistic}}
\end{figure}

In Fig. \ref{fig:ballistic} we simulate the QST of CCA system with
length $N=8$ at the ballistic regime, $J_1=J_7=0.3$, depicted by solid
lines. The dashed lines are the corresponding QST of the system with
uniform coupling strength, $J_n=1$ with the same initial states. It
shows that at time $\tau = 20.7$ the system at ballistic regime get
fidelity larger than $0.99$, while the uniform coupling system get
some mediocre fidelity.


\paragraph*{Experimental Realization.}---
Our proposal can be realized using the experimental realization
mentioned in ref. \cite{Hartmann2006} without atoms in the
cavities. Toroidal micro-cavities can be produced with high precision
and in large number on a chip. These cavities have a very high
Q-factor ($>10^8$) for light that is trapped as whispering gallery
modes and are coupled via tapered optical fibers
\cite{Armani2003}. Another promising candidate for an experimental
realization is photonic crystals \cite{Yariv:99, Akahane2003}. The
technology preparing the coherent or Fock photon state in the cavity
and counting the photon number \cite{Kuhr2007, Wang2008, Brune2008,
  Johnson2010a, Sayrin2011} can be used to compute the fidelity of
the QST of photon state.

\paragraph*{Conclusion.}--- In summary, we propose a QST scheme using
a CCA system to transfer any single-mode photon state from one end of
the array to the opposite end. Our analysis shows that all the results
of QST schemes for spin chain system are applicable in our scheme and
that pretty good QST of any single-mode photon state along the CCA
system can be achieved for arbitrary initial states. Generally there
will be a phase difference between the basis with different photon
number. We eliminate this phase difference by choose the proper cavity
mode frequency $\omega$ depending on transfer time $\tau$. We
numerically simulate the schemes in three cases: uniform coupling CCA,
perfect modulated CCA and the CCA with coupling strengths in ballistic
regime. In every case we use different initial states and sent states,
and the expected results are got. Using the technology of producing
high quality cavity array and precisely preparing and measuring photon
state in a cavity, our scheme of QST along a CCA may be realized in
the near future. 

\begin{acknowledgments}
  This work is supported by NSF of China (Grant No. 11175247) and
  NKBRSF of China (Grant Nos.  2012CB922104 and 2014CB921202).
\end{acknowledgments}

\bibliography{XX_information_flux}

\end{document}